\documentclass[twocolumn,showpacs,amsmath,amssymb,aps,pra,floatfix]{revtex4}
\usepackage{bm}
\usepackage{amsfonts}
\usepackage{graphicx}
\usepackage{graphicx}
\begin{document}

\title{Infrared self-consistent solutions of spinor QED$_3$}
\author{Tomasz Rado\.zycki}
\email{t.radozycki@uksw.edu.pl} \affiliation{Faculty of Mathematics and Natural Sciences, College of Sciences,
Cardinal Stefan Wyszy\'nski University, W\'oycickiego 1/3, 01-938 Warsaw, Poland}

\begin{abstract}
Three-dimensional electrodynamics in the spinor (i.e. two-component) version is considered. With the use of the so called Salam's vertex, the infinite hierarchy of Dyson-Schwinger equations is turned into a set of four self-consistent equations for four parameters describing the infrared behavior of fermion and boson propagators. It is shown numerically, that this set of equations has solutions, at least for certain values of gauge parameter. For weak coupling (i.e. for heavy fermions) the values of all these quantities are found analytically. In the case of massless bare fermion, masses of both particles are generated leading thereby to parity breaking.
\end{abstract}
\pacs{11.10.Kk, 11.15.Tk} 
\maketitle

\section{Introduction}

Studying gauge field theories in low-dimensional space-time allowed for investigations both perturbative and nonperturbative aspects of quantum field theory. Among abelian models of this kind the particular mention deserve quantum electrodynamics in two dimensions --- the so called Schwinger Model (SM)~\cite{schw}) --- and in three dimensions. The former, due to its full solvability, at least in the massless case, has become particularly fruitful and many nonperturbative effects have been discovered and analyzed. On can mention here the generation of the gauge boson mass, via the so called {\em Schwinger mechanism}, fermion confinement, chiral symmetry breaking, the presence of instantons and the $\theta$-vacuum (see for instance~\cite{cks,cadam1,maie,rot,gmc}), bound state formation~\cite{trsing} but also the status of the perturbation series~\cite{sw,os,trpert}. Due to these nontrivial features Schwinger Model has become a testing laboratory for various aspects of the theory of strong interactions. The massive version of this model in no longer solvable but is still attractive thanks to its relative simplicity and remarkable physical content~\cite{cjs,ad1}.

Quantum electrodynamics in three dimensions - the so called Planar QED --- plays an important role in understanding some aspects of QFT too. One can recall here confinement~\cite{bpr,maris,brcr,sauli,srb}, chiral symmetry breaking~\cite{abcw,abkw,brcr,srb,kog}, bound states~\cite{ab,matna,hosh} and the analysis of the gauge dependence of the nonperturbative results with various approximations~\cite{bhr,abar}. Contrary to the SM, the photon becomes in QED$_3$ a dynamic particle and corresponds to real degree of freedom, although it has some peculiar features connected with the fact that the magnetic field is now a (pseudo) scalar and electric one a two-component vector restricted by the Gauss law. The number of dimensions attributes to this model also other specific features. The model may be formulated in two inequivalent versions: the fundamental fermion field may be chosen as four- or two-component one. In the latter --- which we concentrate on in the present work --- the most interesting is the appearance of the topological photon mass term~\cite{djt}. This term may be, just in three dimensions, added to the standard Lagrangian from the very beginning, since it exhibits both Lorentz and gauge (up to the surface terms) invariance --- such a theory is called Chern-Simons QED$_3$ --- but even if absent, it is generated by the interactions with fermions. 

Both low-dimensional models find their applications also beyond QFT: for instance in condense matter physics~\cite{tsvelik,car,lyk}.

Many of the above mentioned phenomena, as confinement, bound states or topological aspects, require for their study nonperturbative language. The problems with perturbation calculations in QED$_3$ manifest themselves also through ambiguity in proper regularization~\cite{djt}.  
The set of Dyson-Schwinger (DS) equations --- which, as one believes, contains the whole information about the quantum system --- constitutes the main tool in nonperturbative investigations of QFT. The common problem in this approach is, that these equations form an infinite hierarchy, the solution of which seems a hopeless task. To deal with this set one is forced to truncate this hierarchy at some level. In two dimensions (in SM), thanks to two gauge symmetries of the theory, it turned out to be possible to express the three-point (vertex) function by fermion propagator (i.e. two-point function) and obtain the close equation for the latter. Not a whole  hierarchy, but just one equation, which turned out to be easily solvable~\cite{tozha,trinst}. Similar method, but now applied to higher functions (five-point), allowed to write a self-consistent equation for two-fermion Green's function~\cite{trjmn}.

In greater number of dimensions, i.e. in QED$_3$ and QED$_4$, gauge symmetry is, however, not sufficient to fully express fermion-photon vertex function by lower ones and therefore we are obliged to assume a certain ansatz for $\Gamma^{\mu}(p+k, p)$. Gauge covariance constitutes here a hint but does not solve the whole problem of truncating the hierarchy. Various such ans\"atze have been proposed based on kinematic properties, renormalizability and requirement of satisfying the Ward-Takahashi identity. One can mention in this context the Ball-Chiu~\cite{bc}, Burden-Roberts~\cite{br} and Curtis-Pennington~\cite{cp} vertices with further improvements~\cite{kiz,brsm}. In the present work, however, we pay particular attention to the vertex introduced by Salam~\cite{sal2} and applied afterwards in the, so called, gauge technique~\cite{del1,dz,tz,park}. The advantage of using Salam's ansatz in the form~(\ref{eq:verp}), apart from its elegance, is that Feynman integrations in DS equations, are similar to those of perturbation calculations for which the computational technique is well elaborated. 

The main idea of the present work is similar to that of our previous paper~\cite{tribb} on QED$_4$, where it was successfully applied and the correct infrared form of propagators was established without the need of infinite renormalization. Here again we assume the possible infrared behavior of propagators up to a set of certain unknown constants and try to determine these constants from the requirement of self-consistency. One has to emphasize that in this approach loop integrations are performed with infrared forms of the Green's functions as integrands. Since we do not wish to extend the results beyond the infrared domain, this procedure is fully acceptable. Firstly we have to note that right hand sides of DS equations~(\ref{eq:dysph}) and~(\ref{eq:dyse}) have the form of convolutions, so the infrared behavior of the integrals should be dictated by the infrared behavior of integrands, secondly our method should not be worse than ordinary perturbation calculations, where {\em free} functions are used as integrands. Finally one has to stress, that Salam's ansatz~(\ref{eq:verp}) in the infrared domain becomes exact.

This work is organized as follows. In Section~\ref{sec:bdi} we collect the main definitions and properties of the model. In Section~\ref{sec:assump} we formulate the assumptions as to the infrared behavior of the photon propagator, fermion propagator and fermion-photon vertex. The following two sections are devoted to obtaining, from the DS equations, the set of four nonlinear equations for four parameters: $\beta$, $\tau$, $\delta m$ and $Z_3$. In the last section we present the numerical results and the dependence of all the parameters on the strength of the interaction. We find also the analytical results for the case of weak coupling.

\section{Basic definitions and identities}
\label{sec:bdi}

Below we briefly summarize conventions used in this work and specify basic properties of Dirac gamma matrices in three dimensions. As to the set of matrices $\gamma^{\mu}$ we use the two-dimensional, i.e. spinor representation in which all gamma's may be expressed by Pauli matrices in the following way:
\begin{eqnarray} 
&&\gamma^0=\sigma_3=\left(\begin{array}{lr}1 & 0 \\ 0 & -1 
\end{array}\right)\; , \;\;\;\;\; 
\gamma^1=i\sigma_1=\left(\begin{array}{lr} 0 & i \\ i & 0 
\end{array}\right)\; , \nonumber\\ 
&&\gamma^2=i\sigma_2=\left( 
\begin{array}{lr} \hspace*{1ex}0 & 1 \\ -1 & 0 \end{array}\right) \; ,
\label{eq:gam}
\end{eqnarray}
and matrix $\gamma^5$ does not exist. Due to this nonexistence the phenomenon of chiral symmetry breaking can be investigated only in four-dimensional representation, unless we introduce even number of flavors in the model.

The metric tensor $g^{\mu\nu}$ and the totally antisymmetric tensor 
$\varepsilon^{\mu\nu\alpha}$ are defined as follows:
\begin{eqnarray}
g^{00} &\!\! = &\!\! -g^{11}=-g^{22}=1\; ,\nonumber\\
 \varepsilon^{012} &\!\! = &\!\! -\varepsilon^{102}=-1\; ,
\label{eq:mette}
\end{eqnarray}
with all other nonzero elements of the latter obtained by cyclic permutations.

In three dimensions the peculiar feature of the choice~(\ref{eq:gam}) is that, apart from ordinary relations:
\begin{eqnarray}
\{\gamma^\mu, \gamma^\nu\}=g^{\mu\nu}\; ,\;\;  \mathrm{tr}\,\gamma^\mu=0\; ,\;\; \mathrm{Tr}\left[\gamma^\mu \gamma^\nu\right]=2g^{\mu\nu}\; ,\nonumber\\
\mathrm{Tr}\left[\gamma^\mu \gamma^\nu \gamma^\rho \gamma^\sigma\right]=2\left(g^{\mu\nu}g^{\rho\sigma}-g^{\mu\rho}g^{\nu\sigma}+g^{\mu\sigma}g^{\nu\rho}\right)\;,
\end{eqnarray} 
also the trace of the product of three (i.e. {\em odd} number) gamma matrices has a nonzero value:
\begin{equation}
\mathrm{Tr}\left[\gamma^\mu \gamma^\nu \gamma^\rho \right]=-2i\varepsilon^{\mu\nu\rho}\; .
\label{eq:proth}
\end{equation}
This identity is (mathematically) responsible for the photon mass generation by the vacuum polarization loop.

In sections~\ref{sec:dsb} and~\ref{sec:dsf} we will also make use of other identities, which can be easily derived:
\begin{eqnarray}
\varepsilon^{\mu\nu\alpha}\gamma_\mu\gamma^\beta\gamma_\nu &\!\! = &\!\! 2\varepsilon^{\mu\beta\alpha}\gamma_\mu+2i\gamma^\alpha\gamma^\beta\; ,\nonumber\\
\varepsilon^{\mu\nu\alpha}\gamma_\mu\gamma_\nu k_\alpha &\!\! = &\!\! -2i \!\!\not \!k \; ,\label{eq:ideg}\\
\gamma^\mu\!\!\not \! k\, \gamma_\mu &\!\! = &\!\! -\!\!\not \! k\; .
\nonumber
\end{eqnarray}

Lagrangian density of three-dimensional electrodynamics  with
gauge fixing term has the following form
\begin{eqnarray} 
{\cal L}(x)=&&\!\!\!\!\!\overline{\Psi}(x)\left(i\gamma^{\mu}\partial_{\mu}-m_0 -
e_0\gamma^{\mu}A_{\mu}(x)\right)\Psi (x)\label{eq:lagr} \\
&&\!\!\!\!\!\!-
\frac{1}{4}F^{\mu\nu}(x)F_{\mu\nu}(x)-
\frac{\lambda}{2}\left(\partial_{\mu}A^{\mu}(x)\right)^2\; , 
\nonumber
\end{eqnarray} 
where $\lambda$ is the gauge parameter. The quantities $m_0$ and $e_0$ are the {\em bare} fermion mass and the {\em bare} coupling constant (charge) respectively and the former may eventually be put equal to zero leading to the parity invariant theory (at least on the classical level). It is worth mentioning that coupling constant has a dimension of $\sqrt{\mathrm{mass}}$ and, as a result, the quantum theory is superrenormalizable. 
The strength tensor $F^{\mu\nu}$ has only three independent elements, two of which ($F^{10}$ and $F^{20}$) constitute to two-component vector of electric field, and one ($F^{12}$) pseudoscalar magnetic field.

It is well known~\cite{djt}, that in three space-time dimensions to this Lagrangian density may be added the so called Chern-Simons (CS) term, proportional to $\varepsilon^{\mu\nu\alpha}A_\mu\partial_\nu A_\alpha$. Although it is not  invariant under gauge transformations, its change reduces to a total derivative inessential for physical observables. Due to this term the photon field acquires a nonzero mass. In this work, however, we do not include explicitly CS term into the Lagrangian and the photon mass will be generated dynamically.

\section{Assumptions for the fundamental Green's functions}
\label{sec:assump}

Underneath the low momenta forms of the dressed two-point Green's functions and three-point vertex are given. They reflect their analytic structures suggested by the perturbation calculation. 

\subsection{Boson propagator}

For the free propagator of massless vector particle we have the usual formula:
\begin{equation}
D^{(0)\mu\nu}(k) = \frac{1}{k^2}\left(-g^{\mu\nu}+
\frac{k^{\mu}k^{\nu}}{k^2}\right) -
\frac{1}{\lambda} \frac{k^{\mu}k^{\nu}}{(k^2)^2}\; .
\label{eq:dprfree}
\end{equation}
It is well known from the perturbative approach~\cite{djt,ch}, that in three space-time dimensions ,,photon'' field acquires a (topological) mass, due to the vacuum polarization process. Even if CS term is initially absent in the Lagrangian, it is generated through the interaction with fermion loop. In the first approximation we can then assume for the dressed propagator
\begin{eqnarray}
D^{\mu\nu}(k) =&&\!\!\!\! \frac{Z_3}{k^2-\tau^2}\left(-g^{\mu\nu}+
\frac{k^{\mu}k^{\nu}}{k^2}\right)\label{eq:dpr}\\
&&\!\!\!+ \frac{iZ_3\tau}{k^2(k^2-\tau^2)}\,\varepsilon^{\mu\nu\rho}k_\rho-
\frac{1}{\lambda} \frac{k^{\mu}k^{\nu}}{(k^2)^2}\; ,
\nonumber
\end{eqnarray}
where $\tau$ is the photon mass to be determined from the consistency conditions, with the renormalization constant $Z_3$ present only in the transverse part of $D^{\mu\nu}(k)$. This is a known fact by virtue of the gauge invariance and the Ward-Takahashi identity:
\begin{equation}
k_\mu D^{\mu\nu}(k)=k_\mu D^{(0)\mu\nu}(k)=-\frac{1}{\lambda}\,\frac{k^\nu}{k^2}\; ,
\label{eq:wtid}
\end{equation}
which ascertains that only transverse part is modified by the interactions.

By the reason of unitarity we expect the nonperturbative value of $Z_3$ to satisfy the condition $0<Z_3\leq 1$~\cite{wei}. In our work on QED$_4$ we obtained this value to be equal to $\frac{7}{9}$~\cite{tribb}. As we will see in section~\ref{sec:rec}, the above bounds will also be satisfied in the present work.

 We treat the denominator $k^2-\tau^2$ in~(\ref{eq:dpr}) as the first term in its Taylor expansion around $k^2=\tau^2$ and potentially admit also higher order polynomial in $k^2$.
 
 The inverse of $D^{\mu\nu}$ is
 \begin{eqnarray}
D^{-1}&&\!\!\!\!\!\!\!  (k)^{\mu\nu}=\label{eq:dinv}\\
&& Z_3^{-1}\bigg[(-k^2g^{\mu\nu}+k^{\mu}k^{\nu})-i\tau\varepsilon^{\mu\nu\rho}k_\rho\bigg]-\lambda k^\mu k^\nu\; .
\nonumber
\end{eqnarray}

\subsection{Fermion propagator}

The free fermion propagator has the standard form:
\begin{equation}
S^{(0)}(p) = \frac{1}{\not\! p - m_0}
\label{eq:sprfree}
\end{equation}
with a pole in the bare mass $m_0$. Contrary to that, the perturbative calculations~\cite{djt} (at least in certain gauges) show that, even if initially massless ,,photon'' acquires a mass $\tau$, the ,,dressed'' fermion propagator should have, as a result of infrared divergences, a branch point at $p^2=m^2$, where $m$ is a physical mass. Therefore we assume for further calculations the following form:
\begin {equation}
S(p) = \frac{1}{(\not\! p - m) (1-p^2/m^2)^{\beta}}\; ,
\label{eq:spr}
\end{equation}
with exponent $\beta$ to be determined (together with mass renormalization constant $\delta m=m-m_0$) by the later requirement of consistency. The additional power $2\beta$ of $p$ in denominator improves the high momentum behavior of loop integrals (after the analytical continuation of~(\ref{eq:spr})), making thereby the theory free of any ultraviolet divergences.

\begin{figure}[h]
\centering
{\includegraphics[width=0.45\textwidth]{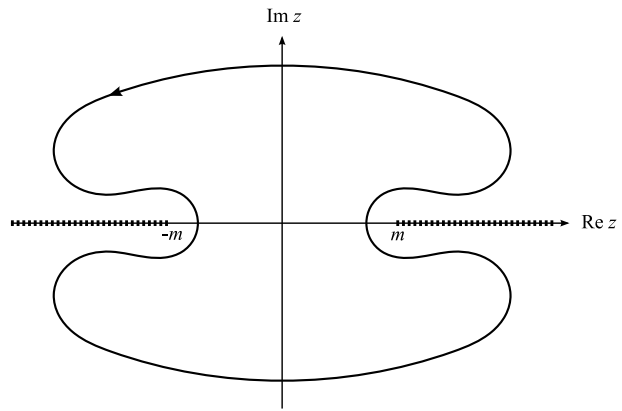}
\caption{The contour of integration in formula~(\ref{eq:dero}), which leads to the spectral function~(\ref{eq:ro}).} \label{fig:contour}}
\end{figure}

To get the spectral representation of~(\ref{eq:spr}), which will be needed for the construction of the vertex, we apply the Cauchy integral formula for the function
$$f(z)=\frac{1}{(m^2-z^2)^\beta}\; ,$$
choosing the contour as shown in Figure~\ref{fig:contour} deformed to sandwich both cuts on the real axis. This is possible for $0<\beta<1$, since otherwise the integrals over the small circles around branch points at $z=\pm m$ diverge. We have then
\begin{eqnarray}
\frac{1}{(m^2-w^2)^\beta} =&&\!\!\!\!\frac{1}{2\pi i}\oint\limits_C \frac{dz}{(m^2-z^2)^\beta(z-w)}\nonumber\\
=&&\!\!\!\! \frac{\sin\pi\beta}{\pi}\bigg[\int\limits_m^\infty \frac{dx}{(x^2-m^2)^\beta (x-w)}\nonumber\\
&&- \int\limits_{-\infty}^{-m} \frac{dx}{(x^2-m^2)^\beta (x-w)}\bigg]\; .
\label{eq:dero}
\end{eqnarray}
Applying this formula for $w=\not\! p$ and taking into account the additional factor $(\not\! p -m)$ in denominator of~(\ref{eq:spr}), we get that way~(c.f.~\cite{tribb})
\begin{eqnarray}
\rho(M) =&&\!\!\!\!\frac{\sin(\pi\beta)}{\pi}\frac{1}{(M-m)
(M^2/m^2-1)^{\beta}}\nonumber\\
&&\!\! \times [\Theta (M-m) - \Theta (-M-m)]\;,
\label{eq:ro}
\end{eqnarray}
where $\Theta$ is the Heaviside step function. We found it useful to
separate out in the above definition of $\rho(M)$ a term, that
describes free propagation of fermion so that $S(p)$ is represented as
\begin{equation}
S(p) = \int dM\rho (M)\left(\frac{1}{\not\! p - m} -
\frac{1}{\not\! p - M}\right)\; .
\label{eq:sp}
\end{equation}

The equations, we obtain in the following sections, turn out to be self-consitent without assuming any nontrivial value of the fermion field renormalization constant $Z_2$ and therefore, in the present work, we have just put it equal to unity. Eventual inclusion of this constant (i.e. an additional unknown variable) would require to generate more equations by adjusting higher Green's functions. This complicates our method, but we keep it in mind and leave for the next paper as a possible following step.

The problem with properly defining constant $Z_2$ raised in~\cite{djt} does not arise in our approach since we do not require the full fermion propagator to have a pole at mass $m$. On the contrary we know, that such a pole does not exist in~(\ref{eq:spr}) due to the infrared divergences and only the additional coefficient in numerator come into play. Its value would be fixed by self-consistency and not by normalization requirement of the residue.

\subsection{Fermion-boson vertex}

Having expressed the full propagator $S$ through the spectral function, we can now make use of the slightly modified (by inclusion of the second term) Salam's ansatz for the vertex function $\Gamma^\mu(p+k, p)$:

\begin{eqnarray}
S(p+k)&&\!\!\!\!\! \Gamma^{\mu}(p+k, p)S(p)\nonumber\\
&&\!\!\!\!\! = \int dM\rho (M)
\Bigg[ \frac{1}{\not\! p\; + \not\! k -m}\gamma^{\mu}\frac{1}
{\not\! p - m}\nonumber\\
&& - \frac{1}{\not\! p\;  + \not\! k -
M}\gamma^{\mu}\frac{1}{\not\! p - M}\Bigg] \; . 
\label{eq:verp}
\end{eqnarray}
By acting on it with a vector $k_\mu$, one automatically obtains the appropriate WT identity, which must be incorporated in any method, that is expected to satisfy gauge invariance:
\begin{widetext}
\begin{eqnarray}
k_\mu S(p+k)\Gamma^{\mu}(p+k, p)S(p) &\!\!\! =&\!\!\!\int dM\rho (M)
\Bigg[ \frac{1}{\not\! p\; + \not\! k -m}\not\! k\frac{1}
{\not\! p - m} - \frac{1}{\not\! p\;  + \not\! k -
M}\not\! k \frac{1}{\not\! p - M}\Bigg]\label{eq:kve}\\
&\!\!\!=&\!\!\!\int dM\rho (M)
\Bigg[ \frac{1}
{\not\! p - m}-\frac{1}{\not\! p\; + \not\! k -m} -  \frac{1}{\not\! p - M}+\frac{1}{\not\! p\;  + \not\! k -
M}\Bigg]=S(p)-S(p+k)\; .
\nonumber
\end{eqnarray}
\end{widetext}
It is a necessary condition but does not guarantee by itself the invariance of physical quantities. 

The longitudinal part of the vertex is then correctly fixed, and transverse part (obviously~(\ref{eq:verp}) is not purely longitudinal) is postulated in an elegant way, naturally being a subject of further improvements~\cite{dz,park}.  

The expressions~(\ref{eq:dinv}), (\ref{eq:sp}) and~(\ref{eq:verp}), together with~(\ref{eq:ro}) will be now required to satisfy DS equations for small momenta.

\section{Dyson-Schwinger equation for the boson propagator}
\label{sec:dsb}

\begin{figure}[h]
\centering
{\includegraphics[width=0.45\textwidth]{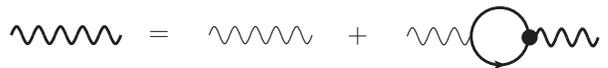}
\caption{Dyson-Schwinger equation for the gauge boson propagator $D^{\mu\nu}(k)$. Light lines represent free propagators and heavy ones dressed propagators. The full circle stands for the full fermion-boson vertex.} \label{fig:dspf}}
\end{figure}

Dyson-Schwinger equation for the photon propagator may be written as

\begin{eqnarray}
&&\!\!\!\!\!D^{\mu\nu}(k)=\frac{1}{k^2}\left(-g^{\mu\alpha}+
\frac{k^{\mu}k^{\alpha}}{k^2}-\frac{1}{\lambda}
\frac{k^{\mu}k^{\alpha}}{k^2}\right)\bigg[\delta_{\alpha}^{\nu}-ie^2_0\times\nonumber\\
&&\!\!\!\!\!\times\mathrm{Tr}\gamma_{\alpha}\int\frac{d^3p}{(2\pi)^3}S(p)\Gamma_{\beta}(p,
p-k)S(p-k)D^{\beta\nu}(k)\bigg],\nonumber\\
\label{eq:dysph}
\end{eqnarray}
and has the graphical representation shown in Figure~\ref{fig:dspf}.

This equation, rewritten for inverse propagator, will be more suitable for deriving self-consistency relations since $D^{\mu\nu}$ decouples from other functions. If we make use of the ansatz~(\ref{eq:verp}), it can be given the form

\begin{eqnarray}
&&\!\!\!\!\!D^{-1}(k)^{\mu\nu}=
-k^2g^{\mu\nu}+k^{\mu}k^{\nu}-\lambda k^{\mu}
k^{\nu }
+ie^2_0\mathrm{Tr}\gamma^{\mu}\times\nonumber\\ 
&&\!\!\times\int dM\rho(M)\int\frac{d^3p}{(2\pi)^3}\Big(\frac{1}
{\not\! p - m + i\varepsilon}\gamma^{\nu}\frac{1}{\not\! p\;  -
\not\! k - m + i\varepsilon} \nonumber\\
&&\!- \frac{1}{\not\! p - M +i
\varepsilon}\gamma^{\nu}\frac{1}{\not\! p\;  - \not\! k - M +
i\varepsilon}\Big)\; . \label{eq:invpp}
\end{eqnarray}

The expression under the integral is well known from one-loop perturbation calculation and its evaluation is straightforward, for instance by passing into euclidean space. It is needless to show any details, so we give only final (Minkowski) result. If we denote 
\begin{eqnarray}
&&\!\!\!\!\!\Pi_m^{\mu\nu}(k)=\label{eq:pimn}\\
&&ie^2_0\mathrm{Tr}\gamma^{\mu}\int\frac{d^3p}{(2\pi)^3}\frac{1}
{\not\! p - m + i\varepsilon}\gamma^{\nu}\frac{1}{\not\! p\;  -
\not\! k - m + i\varepsilon}\; ,
\nonumber
\end{eqnarray}
and introduce Feynman parameters, we find
\begin{eqnarray}
&&\!\!\!\!\!\Pi_m^{\mu\nu}(k)=\nonumber\\
&&\!\!\!\frac{e_0^2}{2\pi}\bigg[(-k^2g^{\mu\nu}+k^{\mu}k^{\nu})\int\limits_0^1 dx\frac{x(1-x)}{(m^2-k^2x(1-x))^{1/2}}\nonumber\\
&&-\frac{i}{2}\, m\varepsilon^{\mu\nu\rho}k_\rho\int\limits_0^1 dx\frac{1}{(m^2-k^2x(1-x))^{1/2}}\; ,
\label{eq:pimnf}
\end{eqnarray}
where $x$ is the Feynman parameter. When $k^2<4m^2$ both integrals are well defined.

The tensor $\Pi_m^{\mu\nu}(k)$ turns out to be transverse with respect to $k_\mu$, as is required by gauge invariance, and particularly by WT identity~(\ref{eq:wtid}). What is important, is the appearance of the novel term, peculiar for three dimensions,  which is proportional to $\varepsilon^{\mu\nu\rho}$ --- a result of the nonzero trace of the product of three gamma matrices~(\ref{eq:proth}). This property is responsible for the gauge field acquiring a mass.

To find the full form of the polarization tensor we have to calculate --- according to~(\ref{eq:verp}) --- the following integral involving the spectral function $\rho(M)$:
\begin{equation}
\Pi^{\mu\nu}(k)=\int dM\rho(M)\left(\Pi_m^{\mu\nu}(k)-\Pi_M^{\mu\nu}(k)\right)\; .
\label{eq:pote}
\end{equation}
Various integrals of that kind, containing spectral function $\rho(M)$, are collected in the appendix~\ref{sec:si}. When we apply  formulae~(\ref{eq:rhoi}) and~(\ref{eq:rhoii}), the only integrals that are left, are those over Feynman parameter $x$:
\begin{eqnarray}
&&\!\!\!\!\!\Pi^{\mu\nu}(k)=\frac{e_0^2\Gamma(\beta+1/2)}{2\pi^{3/2}\Gamma(\beta+1)m}\times\nonumber\\
&&\times\bigg[\left(-k^2g^{\mu\nu}+k^\mu k^\nu\right)\int\limits_0^1 dx\frac{x(1-x)}{1-k^2/m^2\, x(1-x))^{\beta+1/2}}\nonumber\\
&&-\frac{i}{2}m\varepsilon^{\mu\nu\rho}k_\rho\int\limits_0^1 dx\frac{1}{1-k^2/m^2\, x(1-x))^{\beta+1/2}}\bigg]
\label{eq:pmn}
\end{eqnarray}

Both turn out to be expressed through the hypergeometric (Gauss) function (see appendix~\ref{sec:pi}, formulae~(\ref{eq:in1}) and~(\ref{eq:in2})). 
These results allow us to bring the whole Dyson-Schwinger equation~(\ref{eq:invpp}) to the form
\begin{eqnarray}
&&\!\!\!\!\!\!D^{-1}(k)^{\mu\nu}=-\lambda k^{\mu}k^{\nu}+
(-k^2g^{\mu\nu}+k^{\mu}k^{\nu})\times\label{eq:dysftot}\\
&&\!\!\!\!\!\times\left[1+\frac{e_0^2\Gamma(\beta+1/2)}{12\pi^{3/2}\Gamma(\beta+1) m}\:{}_2F_1(2,\beta+1/2;5/2;k^2/4m^2)\right]\nonumber\\
&&\!\!\!\!\!-i\varepsilon^{\mu\nu\rho}k_\rho\frac{e_0^2\Gamma(\beta+1/2)}{4\pi^{3/2}\Gamma(\beta+1)}\:{}_2F_1(1,\beta+1/2;3/2;k^2/4m^2)\nonumber\\
\end{eqnarray}
and adjust the first two terms in the infrared domain, i.e. close to the photon mass ($k^2\approx\tau^2$). In that way we get two equations for $Z_3$ and $\tau$:
\begin{eqnarray}
Z_3^{-1} =&&\!\!\!\!\! 1+\frac{e_0^2\Gamma(\beta+1/2)}{12\pi^{3/2} m\Gamma(\beta+1)}\times\nonumber\\
&&\times\:{}_2F_1(2,\beta+1/2;5/2;\tau^2/4m^2)\; ,\label{eq:eq1}\\
\tau Z_3^{-1} =&&\!\!\!\!\! \frac{e_0^2\Gamma(\beta+1/2)}{4\pi^{3/2} m\Gamma(\beta+1)}\times\nonumber\\
&&\times\:{}_2F_1(1,\beta+1/2;3/2;\tau^2/4m^2)\; ,
\label{eq:eq2}
\end{eqnarray}
which will be solved in section~\ref{sec:rec} together with the other two derived form the DS equation for the fermion propagator.

\section{Dyson-Schwinger equation for the fermion propagator}
\label{sec:dsf}

\begin{figure}[h]
\centering
{\includegraphics[width=0.45\textwidth]{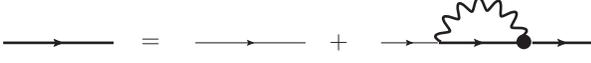}
\caption{Dyson-Schwinger equation for the fermion propagator. As in Figure~\ref{fig:dspf}, heavy lines stand for full functions and light for free ones.} \label{fig:dsef}}
\end{figure}

Dyson-Schwinger equation for the fermion propagator has the form
\begin{eqnarray}
S(p)=&&\!\!\!\!\! 
\frac{1}{\not\! p - m_0}\bigg[1+ie^2_0\gamma^{\mu}\times\label{eq:dyse}\\
&&\!\!\!\!\!\times\int\frac{d^3k}{(2\pi)^3}S(p+k)\Gamma^{\nu}(p
+k, p)S(p)D_{\mu\nu}(k)\bigg]\; ,
\nonumber
\end{eqnarray}
and may be represented as shown in Figure~\ref{fig:dsef}.

It can be rewritten in a simpler manner
\begin{equation}
(\not\! p -m_0)S(p)=1+\Sigma(p)S(p)\; ,
\label{eq:hs}
\end{equation}
where
\begin{eqnarray}
&&\!\!\!\!\!\Sigma(p)S(p)=\label{eq:sis}\\
&&=ie^2_0\gamma^{\mu}\int\frac{d^3k}{(2\pi)^3}S(p+k)
\Gamma^{\nu}(p+k, p)S(p)D_{\mu\nu}(k)\; .
\nonumber
\end{eqnarray}

Now we have to substitute (\ref{eq:verp}) for the vertex, and~(\ref{eq:dpr}) for $D^{\mu\nu}$, obtaining
\begin{eqnarray}
\Sigma(p)S(p)=&&\!\!\!\!\! [\Sigma(p)S(p)]_A+[\Sigma(p)S(p)]_B\nonumber\\
&&\!\!\!\!\! +[\Sigma(p)S(p)]_C+[\Sigma(p)S(p)]_D\; ,
\label{eq:sis1}
\end{eqnarray}
where, to avoid lengthy expressions, we separated the contributions coming from different tensor structures in the photon propgator: $g^{\mu\nu}$, $k^\mu k^\nu$, $\varepsilon^{\mu\nu\rho}k_\rho$, and from gauge-dependent longitudinal part:

\begin{widetext}
\begin{eqnarray}
&&\!\!\!\!\![
\Sigma (p)S(p)]_A=\label{eq:siga}\\ 
&&-iZ_3e^2_0\int
dM\rho(M)\int\frac{d^3k}{(2\pi)^3}\left[ \gamma^{\mu}\frac{1}{\not\! p\; + \not\! k -
m+i\varepsilon}\gamma_{\mu}\frac{1}{(k^2-\tau^2+i\varepsilon)(\not\! p - m+i\varepsilon)} - (m\rightarrow M)\right]\nonumber\\
&&\!\!\!\!\![\Sigma (p)S(p)]_B=\label{eq:sigb}\\
&& iZ_3e^2_0\int dM\rho(M)\int\frac{d^3k}{(2\pi)^3}
\left[ \not\! k\frac{1}{\not\! p\; + \not\! k - m
+i\varepsilon}\not\! k\frac{1}{(k^2+i\varepsilon)(k^2-\tau^2+i\varepsilon)(\not\! p - m+i\varepsilon)}- (m\rightarrow M)\right]\nonumber\\
&&\!\!\!\!\![\Sigma (p)S(p)]_C=\label{eq:sigc}\\ 
&& -Z_3e^2_0\tau\int
dM\rho(M)\int\frac{d^3k}{(2\pi)^3}\left[ \varepsilon^{\mu\nu\rho}\gamma_{\mu}\frac{1}{\not\! p\; + \not\! k -
m+i\varepsilon}\gamma_{\nu}\frac{k_\rho}{(k^2+i\varepsilon)(k^2-\tau^2+i\varepsilon)(\not\! p - m+i\varepsilon)}- (m\rightarrow M)\right]\nonumber\\
&&\!\!\!\!\![\Sigma (p)S(p)]_D
= \label{eq:sigd}\\
&&-\frac{ie^2_0}{\lambda}\int dM\rho(M)\int\frac{d^3k}{(2\pi)^3}
\left[ \not\! k\frac{1}{\not\! p\; + \not\! k - m
+i\varepsilon}\not\! k\frac{1}{(k^2+i\varepsilon)^2(\not\! p - m+i\varepsilon)} - (m\rightarrow M)\right]\nonumber\; .
\end{eqnarray}

Each of the above momentum integrals can be performed in an ordinary way by performing Wick's rotation and introducing Feynman parameters. In the symmetric integration no infinities arise even if the simple power counting indicate an apparent logarithmic divergence. Disregarding all details of the computation, as a result we have (again in Minkowski space)
\begin{eqnarray}
I_A=&&\!\!\!\!\!ie^2_0\int\frac{d^3k}{(2\pi)^3}\gamma^{\mu}\frac{1}{\not\! p\; + \not\! k -
m+i\varepsilon}\gamma_{\mu}\frac{1}{k^2-\tau^2+i\varepsilon}=\frac{e_0^2}{8\pi}\int\limits_0^1dx\frac{\not\! p(1-x)-3m}{(m^2x-p^2x(1-x)+\tau^2(1-x))^{1/2}}\; ,
\label{eq:i1}\\
I_B=&&\!\!\!\!\!ie^2_0\int\frac{d^3k}{(2\pi)^3}
 \not\! k\frac{1}{\not\! p\; + \not\! k - m
+i\varepsilon}\not\! k\frac{1}{(k^2+i\varepsilon)(k^2-\tau^2+i\varepsilon)}=\frac{e_0^2}{8\pi}\int\limits_0^1dx\bigg[\frac{1}{(m^2x-p^2x(1-x)+\tau^2(1-x))^{1/2}}\nonumber\\
&&+\frac{\not\! p(\not\! p+m)}{\tau^2}\left(\frac{x}{(m^2x-p^2x(1-x))^{1/2}}-\frac{x}{(m^2x-p^2x(1-x)+\tau^2(1-x))^{1/2}}\right)\bigg](\not\! p -m)\; ,
\label{eq:i2}\\
I_C=&&\!\!\!\!\!e^2_0\int\frac{d^3k}{(2\pi)^3}\varepsilon^{\mu\nu\rho}\gamma_{\mu}\frac{1}{\not\! p\; + \not\! k -
m+i\varepsilon}\gamma_{\nu}\frac{k_\rho}{(k^2+i\varepsilon)(k^2-\tau^2+i\varepsilon)}=\frac{e_0^2}{4\pi }\int\limits_0^1dx\bigg[\frac{-1}{(m^2x-p^2x(1-x)+\tau^2(1-x))^{1/2}}\nonumber\\
&&+\frac{\not\! p(\not\! p-m)}{\tau^2}\left(\frac{x}{(m^2x-p^2x(1-x)+\tau^2(1-x))^{1/2}}-\frac{x}{(m^2x-p^2x(1-x))^{1/2}}\right)\bigg]\; ,
\label{eq:i3}\\
I_D=&&\!\!\!\!\! ie^2_0\int\frac{d^3k}{(2\pi)^3}
 \not\! k\frac{1}{\not\! p\; + \not\! k - m
+i\varepsilon}\not\! k\frac{1}{(k^2+i\varepsilon)^2}=\frac{e_0^2}{8\pi}\int\limits_0^1 dx\bigg[\frac{1}{(m^2x-p^2x(1-x))^{1/2}}\nonumber\\&&
+\frac{\not \! p(\not\! p+m)}{2}\frac{x(1-x)}{(m^2x-p^2x(1-x))^{3/2}}\bigg](\not\! p -m)\; .
\label{eq:i4}
\end{eqnarray}
\end{widetext}

Now the contributions to $\Sigma(p)S(p)$ may be written as
\begin{eqnarray}
&&\!\!\!\!\! [\Sigma (p)S(p)]_A =\label{eq:ssa}\\
&&-Z_3\int dM\rho(M)\left[I_A\frac{1}{\not\! p-m+i\varepsilon}-(m\rightarrow M)\right]\; ,\nonumber\\
&&\!\!\!\!\! [\Sigma (p)S(p)]_B =\label{eq:ssb}\\
&&Z_3\int dM\rho(M)\left[I_B\frac{1}{\not\! p-m+i\varepsilon}-(m\rightarrow M)\right]\; ,\nonumber\\
&&\!\!\!\!\! [\Sigma (p)S(p)]_C =\label{eq:ssc}\\
&&-Z_3\tau\int dM\rho(M)\left[I_C\frac{1}{\not\! p-m+i\varepsilon}-(m\rightarrow M)\right]\; ,\nonumber\\
&&\!\!\!\!\! [\Sigma (p)S(p)]_D =\label{eq:ssd}\\
&&-\frac{1}{\lambda}\int dM\rho(M)\left[I_D\frac{1}{\not\! p-m+i\varepsilon}-(m\rightarrow M)\right]\; ,\nonumber
\end{eqnarray}
and the appropriate spectral integrals, after a small rearrangement, are carried out according to formulae of the appendix~\ref{sec:si}. 

Before performing $x$ integrations, it is important to notice, that if one substitutes~(\ref{eq:spr}) into the left hand side of~(\ref{eq:dyse}) and multiply both sides by $(\not\! p +m_0)$, there appear two singular terms, when $p^2\rightarrow m^2$:
\begin{equation}
-\frac{m^{2\beta}\delta m (\not\! p +m)}{(m^2-p^2)^{\beta+1}}+\frac{m^{2\beta}}{(m^2-p^2)^{\beta}}\; .
\label{eq:singl}
\end{equation}
For our goal it is sufficient to pick out the identical terms from the right hand side of~(\ref{eq:dyse}). It is a very pleasant observation that it actually displays identical analytical structure, which is a strong sign of self-consistency. Consequently we do not need to preform explicitly all parameter integrations, and can limit ourselves to those terms that contain singularities of the kind~(\ref{eq:singl}). The appropriate results are given below:
\begin{eqnarray}
&&\!\!\!\!\! [\Sigma (p)S(p)]_A \approx -\frac{e_0^2Z_3}{8\pi}{\cal K}_1(p^2,\tau^2)\times\label{eq:ssa1}\\
&&\times\frac{\not\! p(\not\! p+m)}{(m^2-p^2)(1-p^2/m^2)^\beta}\; ,\nonumber
\\
&&\!\!\!\!\! [\Sigma (p)S(p)]_B \approx 0\label{eq:ssb1}\\
&&\!\!\!\!\! [\Sigma (p)S(p)]_C \approx -\frac{e_0^2Z_3\tau}{4\pi}{\cal K}_2(p^2,\tau^2)\times\label{eq:ssc1}\\
&&\times\frac{\not\! p+m}{(m^2-p^2)(1-p^2/m^2)^\beta}\; ,\nonumber\\
&&\!\!\!\!\! [\Sigma (p)S(p)]_D \approx \frac{e_0^2}{8\pi\lambda m}\bigg(\frac{-1}{(m^2-p^2)(1-p^2/m^2)^\beta}\nonumber\\
&&+\frac{1-\beta}{2\beta}\frac{1}{m^2(1-p^2/m^2)^\beta}\bigg)\not\! p(\not\! p+m)\label{eq:ssd1}\; ,
\end{eqnarray}
where $\approx 0$ means that the expression does not diverge, when $p^2\rightarrow m^2$. The functions ${\cal K}_1$ and ${\cal K}_2$ have the following form:
\begin{eqnarray}
{\cal K}_1(p^2,\tau^2)&\!\!\! =&\!\!\!\int\limits_0^1 dx\frac{x+2}{(p^2x^2+\tau^2(1-x))^{1/2}}\; ,
\label{eq:k1}\\
{\cal K}_2(p^2,\tau^2)&\!\!\! =&\!\!\!\int\limits_0^1 dx\frac{1}{(p^2x^2+\tau^2(1-x))^{1/2}}\; .
\label{eq:k2}
\end{eqnarray}

Equating identical divergent terms on both sides of Dyson-Schwinger equation~(\ref{eq:hs}) we obtain two relations for parameters $\delta m$ and $\beta$:
\begin{eqnarray}
\delta m &\!\!\! =&\!\!\! \frac{e_0^2 Z_3}{8\pi}\left({\cal S}_1(m,\tau)+\frac{1}{\lambda Z_3}\right)\; ,\label{eq:eq3}\\
1 &\!\!\! =&\!\!\! \frac{e_0^2 Z_3}{8\pi}\left({\cal S}_2(m,\tau)+\frac{1}{\lambda Z_3 m\beta}\right)\; ,\label{eq:eq4}
\end{eqnarray}
where auxiliary quantities ${\cal S}_1(m,\tau)$ and ${\cal S}_2(m,\tau)$ are defined as
\begin{eqnarray}
{\cal S}_1&\!\!\!=&\!\!\! m{\cal K}_1(m^2,\tau^2)+2\tau{\cal K}_2(m^2,\tau^2)\label{eq:s12a}\\
{\cal S}_2&\!\!\!=&\!\!\!{\cal K}_1(m^2,\tau^2)+2m^2\frac{\partial {\cal K}_1(m^2,\tau^2)}{\partial m^2}+4m\tau\frac{\partial {\cal K}_2(m^2,\tau^2)}{\partial m^2} ,\nonumber\\
\label{eq:s12b}
\end{eqnarray}
and found in appendix~\ref{sec:pi}.

\section{Results and conclusions}
\label{sec:rec}

Below we rewrite the whole set of equations using {\em renormalized} quantities: fermion mass $m$, gauge coupling constant $e=Z_3^{1/2} e_0$ and gauge parameter $\lambda_R=Z_3\lambda$. We additionally introduce a dimensionless parameter $\zeta=\frac{e^2}{4\pi m}$, obtaining
\begin{eqnarray}
\frac{\delta m}{m} = &&\!\!\!\!\!\frac{\zeta}{2}\bigg[1-\frac{\tau}{m}+\left(2+\frac{2\tau}{m}+\frac{\tau^2}{2m^2}\right)\times\nonumber\\ 
&&\times\ln (2m/\tau+1)+\frac{1}{\lambda_R}\bigg]\; ,\label{eq:deleq}\\
1 = &&\!\!\!\!\!\frac{\zeta}{2}\bigg[2+\frac{2\tau}{m}-\left(\frac{2\tau}{m}+\frac{\tau^2}{m^2}\right)\times\nonumber\\ 
&&\times\ln (2m/\tau+1)+\frac{1}{\lambda_R\beta}\bigg]\; ,\label{eq:jedeq}\\
Z_3 = &&\!\!\!\!\! 1-\frac{\zeta}{3}\frac{\Gamma(\beta+1/2)}{\sqrt{\pi}\Gamma(\beta+1)}\:{}_2F_1(2,\beta+1/2;5/2;\tau^2/4m^2)\; ,\nonumber\\
\label{eq:z3eq}\\
\frac{\tau}{m} = &&\!\!\!\!\! \zeta\frac{\Gamma(\beta+1/2)}{\sqrt{\pi}\Gamma(\beta+1)}\:{}_2F_1(1,\beta+1/2;3/2;\tau^2/4m^2)\; .\label{eq:taueq}
\end{eqnarray} 

Now the question arises whether this set of highly nonlinear equations for unknown $\delta m/m$, $\beta$, $Z_3$ and $\tau/m$ has a certain domain of solvability. The answer is positive for a large range of values of gauge parameter $\lambda_R$ and for weak coupling. For exemplary value $\lambda_R=1$ numerically found solutions of all equations as functions of the parameter $\zeta$ are presented below. 

In the first graph of Figure~\ref{fig:graph1} we show the dependence of the topological photon mass $\tau$ (in units of $m$) on the parameter $\zeta$. One can see, that for small values of parameter it approaches the value $e^2/4\pi$~\cite{ch}. In spite of initial zero value of the photon mass, it reappears as an outcome of interactions with fermions.

On the second graph the dependence of power $\beta$ on $\zeta$ is presented. As we remember our method is reliable only for $0<\beta<1$. This corresponds roughly to $0<\zeta<1$. 

\begin{figure}[h]
\centering
{\includegraphics[width=0.49\textwidth]{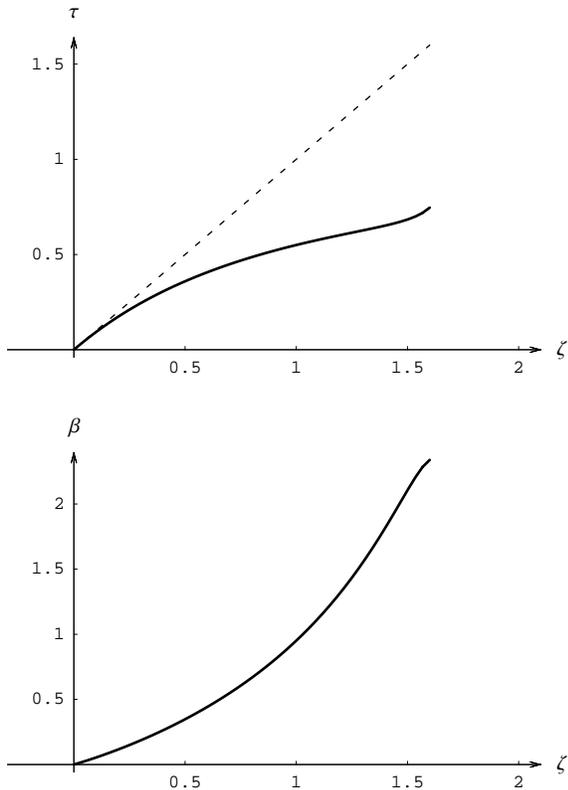}
\caption{The dependence of photon mass $\tau$ in units of $m$ (upper plot) and power $\beta$ (lower plot) on the parameter $\zeta$. The dashed line corresponds to $\tau/m=\zeta$.} \label{fig:graph1}}
\end{figure}

\begin{figure}[h]
\centering
{\includegraphics[width=0.49\textwidth]{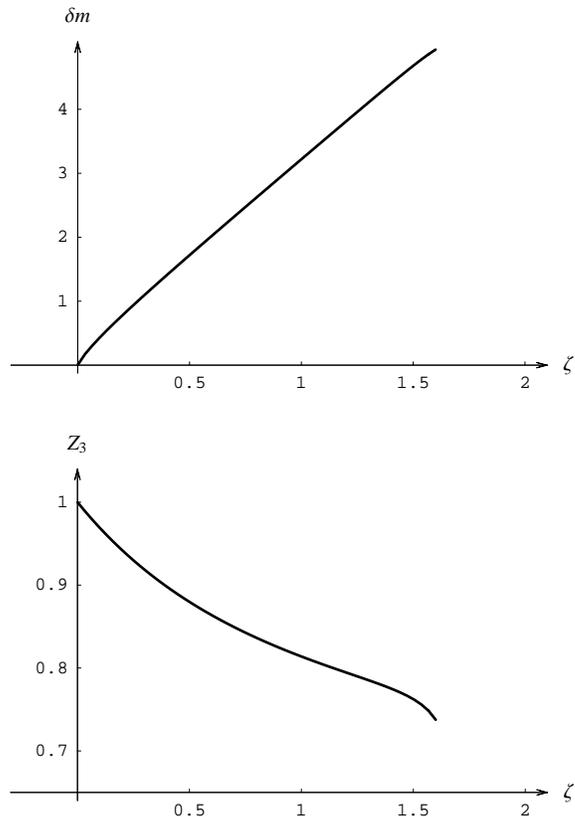}
\caption{The dependence of mass renormalization $\delta m$ in units of $m$ (upper plot) and charge renormalization constant $Z_3$ (lower plot) on the parameter $\zeta$.} \label{fig:graph2}}
\end{figure}

Figure~\ref{fig:graph2} shows the fermion mass renormalization $\delta m$ (again in units of $m$) and coupling constant renormalization constant $Z_3$ as functions of $\zeta$. Please note, that $\delta m<m$ (for small $\zeta$) and $0<Z_3<1$ and it decreases with increasing coupling.

The comparison of all numerical results suggests that it might be possible to find analytical solutions in the case of weak coupling (which corresponds to heavy fermion), i.e. when $\zeta\ll 1$. From the Figure~\ref{fig:graph1} we conclude that in such a case we should assume $\tau/m\ll 1$. We suppose also that $\beta\ll 1$. All these assumptions will be justified {\em a posteriori}, when the whole set of parameters is determined. 

For small values of $\tau$ the equation~(\ref{eq:jedeq}) may be given the approximate form (considering gauges, where $\xi_R$ is of order of unity)
\begin{equation}
1 = \frac{\zeta}{2}\,\frac{1}{\lambda_R\beta}\; ,\label{eq:jedeq1}
\end{equation}
since all other terms, particularly those containing a quotient $\tau/m$ are small compared to $1/\beta$ and have been neglected. Consequently we have
\begin{equation}
\beta=\frac{\zeta}{2\lambda_R}\; ,
\label{eq:betas}
\end{equation}
which is in agreement with the assumption $\beta\ll 1$. 

For small values of $\tau$ and $\beta$ the equation~(\ref{eq:taueq}) reduces to (recall that $\Gamma(1/2)=\sqrt{\pi}$ and $\:{}_2F_1(a,b;c;0)=1$)
\begin{equation}
\frac{\tau}{m}=\zeta\; ,
\label{eq:taus}
\end{equation}
which is the Coleman's and Hill's result~\cite{ch}. The first plot of Figure~\ref{fig:graph1} shows, however, that for larger values of $\zeta$ it deviates form the straight line. This is well understood since in Coleman's and Hill's paper the value of the photon mass was defined as $\tau^2=\Pi(0)$, where $\Pi$ is a polarization scalar. In our work the position of the photon mass is determined by the equation $\tau^2=\Pi(\tau^2)$. One should also keep in mind that our solution is only approximate.

From~(\ref{eq:taus}) we find again that the assumption $\tau/m\ll 1$ is justified. In a similar manner we get the following result for $Z_3$:
\begin{equation}
0<Z_3=1-\frac{\zeta}{3}<1\; ,
\label{eq:z3s}
\end{equation}
and for $\delta m$:
\begin{equation}
\frac{\delta m}{m}=\frac{\zeta}{2}\left( 1-2\ln(\zeta/2)+\frac{1}{\lambda_R}\right)\; .
\label{eq:dms}
\end{equation}
The unpleasant feature is the gauge dependence of the mass renormalization, which might indicate the similar gauge dependence of the physical fermion mass. It is, however, a common feature of perturbative and nonperturbative results in QED$_3$~\cite{moroz,mnu}.

In the weak coupling regime one can obtain also the analytical results for the Landau gauge, which corresponds to taking the limit $\lambda_R\rightarrow \infty$. As is known, in this case the infrared singularities disappear~\cite{djt,moroz} so we expect $\beta\rightarrow 0$, since the fermion propagator should have an ordinary pole at physical mass and not a branch point. Therefore, before taking the limit, we put $\beta=\beta_0/\lambda_R$ with constant $\beta_0$ to be determined. With all other assumptions identical as above, we get similar results as before: $\tau/m=\zeta$, $Z_3=1-\zeta/3$ and
\begin{equation}
1=\frac{\zeta}{2}\left(2+\frac{1}{\beta_0}\right)\;\; \Longrightarrow \;\; \beta_0=\frac{\zeta}{2(1-\zeta)}\; ,
\label{eq:zela}
\end{equation}
and hence for $\zeta\ll 1$
\begin{equation}
\beta=\frac{\zeta}{2\lambda_R(1-\zeta)}\approx \frac{\zeta}{2\lambda_R}\;\;\underset{\lambda_R\rightarrow \infty}{\longrightarrow} 0\; .
\label{eq:zelab}
\end{equation}
The result for the mass renormalization in this case is
\begin{equation}
\frac{\delta m}{m}=\frac{\zeta}{2}\left( 1-2\ln(\zeta/2)\right)\; .
\label{eq:dmsl}
\end{equation}

The nonanalyticity in $\zeta$ observed in the last formula and in~(\ref{eq:dms}), which manifests itself through the presence of a logarithmic function of the coupling constant, is similar in nature to the one found by us in QED$_4$~\cite{tribb}, where it constituted a reflection of the possible ill behavior of the perturbation series ~\cite{dyson}.

\begin{figure}[t]
\centering
{\includegraphics[width=0.49\textwidth]{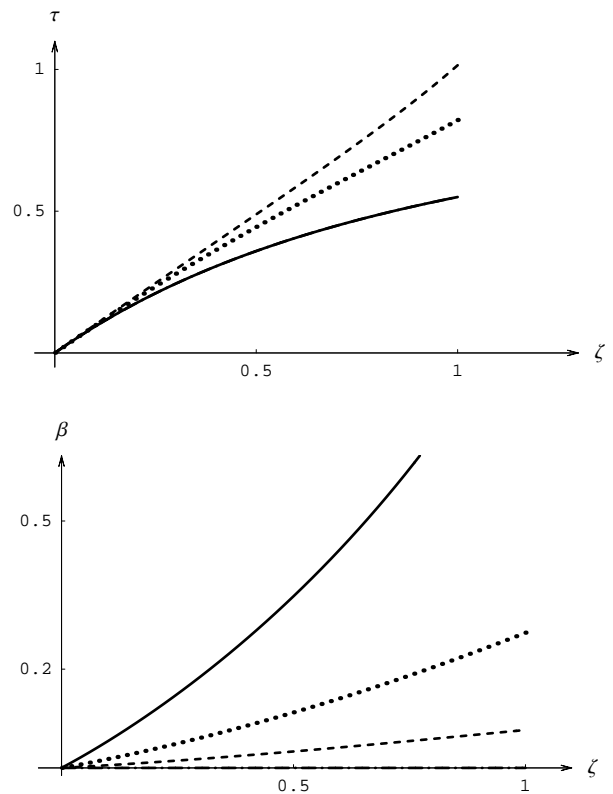}
\caption{The comparison of the behavior  of photon mass $\tau$ (upper plot) and parameter $\beta$ (lower plot) for different gauges: solid line -- $\lambda_R=1$, dotted line -- $\lambda_R=3$, dashed line -- $\lambda_R=10$, mixed line -- $\lambda_R=\infty$ (Landau gauge). The latter is not visible on the upper plot, since it is overlapping with the solid line. } \label{fig:graph3}}
\end{figure}

\begin{figure}[h]
\centering
{\includegraphics[width=0.49\textwidth]{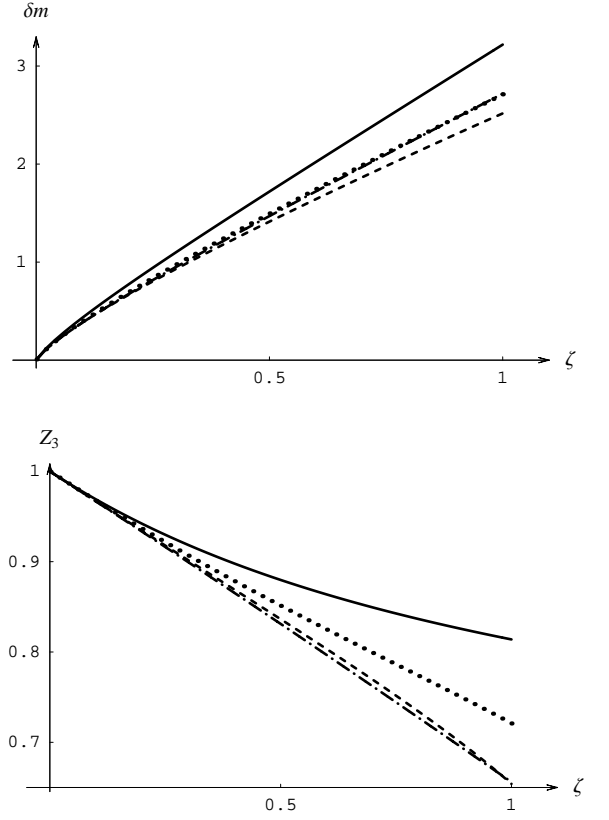}
\caption{The comparison of the behavior of fermion mass renormalization $\delta m$ (upper plot) and charge renormalization constant $Z_3$ (lower plot) for different gauges: solid line -- $\lambda_R=1$, dotted line -- $\lambda_R=3$, dashed line -- $\lambda_R=10$, mixed line -- $\lambda_R=\infty$ (Landau gauge). On the upper plot the lines for $\lambda_R=3$ and for Landau gauge follow close curves.} \label{fig:graph4}}
\end{figure}

The solutions of the set of nonlinear equations~(\ref{eq:deleq}--\ref{eq:taueq}) involving gauge parameter $\lambda_R$ can, in general, be gauge dependent. To clarify this point we performed a couple of plots of the model parameters for different gauges, i.e. for  $\lambda_R=1,3,10,\infty$, the last case corresponding to the Landau gauge. In Figure~\ref{fig:graph3} the dependence of $\tau$ and $\beta$ on $\lambda_R$ is demonstrated. Figure~\ref{fig:graph4} shows the same dependence of $\delta m$ and $Z_3$.

It may be easily seen, that, apart from the parameter $\beta$, which is and may be obviously gauge dependent, all other values become practically gauge independent for weak coupling. To get the reliable results for strong coupling we need to go beyond the first approximation consisting of the simple assumptions~(\ref{eq:dpr}),~(\ref{eq:spr}) and~(\ref{eq:verp}). The same refers to other nonperturbative (but not exact) calculations.

At the end we would like to note, that from the dependence of $\delta m$ (Figure~\ref{fig:graph2} and Figure~\ref{fig:graph4}) one sees, that for certain values of parameter $\zeta$ we have $\delta m = m$ (i.e. $m_0=0$). Assuming value $\delta m/m=1$ as fixed, the equations~(\ref{eq:deleq})--(\ref{eq:taueq}) may be in turn solved for $\zeta$, $\tau$, $\beta$ and $Z_3$. In the case of the Landau gauge the following values are obtained
\begin{equation}
\frac{\tau}{m}\approx\zeta\approx 0.34\; ,
\label{eq:numtz}
\end{equation}
from which we get the generated masses (the Lagrangian in this case does not contain any masses) to be
\begin{equation}
m\approx 2.94\times \frac{e^2}{4\pi}\; ,\;\;\;\; \tau=\frac{e^2}{4\pi}\; ,
\label{eq:nummt}
\end{equation}
and the parity becomes broken. For two other parameters we find
\begin{equation}
\beta_0\approx 0.21\;\; (\mathrm{naturally}\;\beta=0)\; ,\;\;\;\; Z_3\approx 0.89\; .
\label{eq:numbz}
\end{equation}

A comment should be made here. The obtained result of parity breaking is not at variant with Vafa-Witten theorem~\cite{vawi,ein} stating, that there is no spontaneous parity breaking in theories with vectorlike fermions. The three -dimensional QED with two-component fermions is very special theory also in that aspect, that it evades the proof of this theorem. The crucial point in the Vafa's and Witten's argument is the positivity of the boson measure (after integrating out the fermion degrees of freedom) which demands the positivity of the determinant of the Dirac operator. This positivity may be easily proved in four component version of the theory, because by the application of the matrix $\gamma^5$ one sees that eigenvalues are always paired in such a way that the unwanted signs cancel. However, in  spinor version of QED no such $\gamma^5$ matrix exists (there is no matrix anticommuting with matrices $\gamma^0$, $\gamma^1$ and $\gamma^2$) so the above argument fails. 

The results of~\cite{tapp}, which support the Vafa-Witten conclusion, are not in conflict with the present work either. They were obtained with the assumptions $m\ll\alpha$ or $m\gg\alpha$, which correspond to $\zeta\gg 1$ or $\zeta\ll 1$ respectively, while in our case we have $\zeta\approx 0.34$. On has to stress also, that the conclusions of~\cite{tapp} are got in the limit of infinitely many flavors, while we have one flavor theory. The number of flavors, particularly if it is even, may be essential for the positivity of the determinant spoken above.

\appendix
\section{Spectral integrals}
\label{sec:si}

Below we collect the set of integrals involving spectral density~(\ref{eq:ro}), that we used in sections~\ref{sec:dsb} and~\ref{sec:dsf}. First one can find that for $a^2<m^2$, one has
\begin{eqnarray}
&&\!\!\!\!\!\!\int dM\rho(M)\left(\frac{1}{(m^2-a^2)^{1/2}}-\frac{1}{(M^2-a^2)^{1/2}}\right)=\nonumber\\
&&\frac{m^{2\beta}\Gamma(\beta+1/2)}{\sqrt{\pi}\Gamma(\beta+1)(m^2-a^2)^{\beta+1/2}}\; ,
\label{eq:rho1}
\end{eqnarray}
and similarly
\begin{eqnarray}
&&\!\!\!\!\!\!\int dM\rho(M)\left(\frac{1}{(m^2-a^2)^{3/2}}-\frac{1}{(M^2-a^2)^{3/2}}\right)=\nonumber\\
&&\frac{2m^{2\beta}\Gamma(\beta+3/2)}{\sqrt{\pi}\Gamma(\beta+1)(m^2-a^2)^{\beta+3/2}}\; ,
\label{eq:rho1a}
\end{eqnarray}
where $\Gamma$ is the Euler function. From the former result by the appropriate substitution we obtain
\begin{eqnarray}
&&\!\!\!\!\!\int dM\rho(M)\left(\frac{1}{(m^2-k^2x(1-x))^{1/2}}-(m\rightarrow M)\right)\nonumber\\
&&=\frac{\Gamma(\beta+1/2)}{\sqrt{\pi} \Gamma(\beta+1)}\frac{m^{2\beta}}{(m^2-k^2 x(1-x))^{\beta+1/2}}\; ,\label{eq:rhoi}\\
&&\!\!\!\!\!\int dM\rho(M)\left(\frac{m}{(m^2-k^2x(1-x))^{1/2}}-(m\rightarrow M)\right)\nonumber\\
&&=\frac{\Gamma(\beta+1/2)}{\sqrt{\pi} \Gamma(\beta+1)}\frac{m^{2\beta+1}}{(m^2-k^2\, x(1-x))^{\beta+1/2}}\; ,
\label{eq:rhoii}
\end{eqnarray}
where we additionally used the relation
\begin{equation}
\int dM\rho(M) M f(M^2)= m\int dM\rho(M)f(M^2)\; .
\label{eq:rofm}
\end{equation}

With a slight modification of~(\ref{eq:rhoi}) and~(\ref{eq:rhoii}) one can easily get the other set of integrals:
\begin{eqnarray}
&&\!\!\!\!\!\int dM\rho(M)\left(\frac{1}{(m^2x-k^2x(1-x))^{1/2}}-(m\rightarrow M)\right)\nonumber\\
&&=\frac{\Gamma(\beta+1/2)}{\sqrt{\pi} \Gamma(\beta+1)}\frac{m^{2\beta}x^\beta}{(m^2 x-k^2\, x(1-x))^{\beta+1/2}}\; ,\label{eq:rhoia}\\
&&\!\!\!\!\!\int dM\rho(M)\left(\frac{m}{(m^2x-k^2x(1-x))^{1/2}}-(m\rightarrow M)\right)\nonumber\\
&&=\frac{\Gamma(\beta+1/2)}{\sqrt{\pi} \Gamma(\beta+1)}\frac{m^{2\beta+1}x^\beta}{(m^2x-k^2\, x(1-x))^{\beta+1/2}}\; ,
\label{eq:rhoiia}
\end{eqnarray}

From~(\ref{eq:rho1a}) one finds
\begin{eqnarray}
&&\!\!\!\!\!\int dM\rho(M)\left(\frac{1}{(m^2x-k^2x(1-x))^{3/2}}-(m\rightarrow M)\right)\nonumber\\
&&=\frac{2\Gamma(\beta+3/2)}{\sqrt{\pi} \Gamma(\beta+1)}\frac{m^{2\beta}x^\beta}{(m^2 x-k^2\, x(1-x))^{\beta+3/2}}\; ,\label{eq:rhoib}\\
&&\!\!\!\!\!\int dM\rho(M)\left(\frac{m}{(m^2x-k^2x(1-x))^{3/2}}-(m\rightarrow M)\right)\nonumber\\
&&=\frac{2\Gamma(\beta+3/2)}{\sqrt{\pi} \Gamma(\beta+1)}\frac{m^{2\beta+1}x^\beta}{(m^2x-k^2\, x(1-x))^{\beta+3/2}}\; .
\label{eq:rhoiib}
\end{eqnarray}
and also
\begin{eqnarray}
&&\!\!\!\!\!\!\int dM\rho(M)\times\nonumber\\
&&\!\times\left(\frac{1}{(m^2x-k^2x(1-x)+\tau^2(1-x))^{1/2}}-(m\rightarrow M)\right)\nonumber\\
&&\!=\frac{\Gamma(\beta+1/2)}{\sqrt{\pi} \Gamma(\beta+1)}\frac{m^{2\beta}x^\beta}{(m^2 x-k^2\, x(1-x)+\tau^2(1-x))^{\beta+1/2}}\, .\nonumber\\
\label{eq:rhoic}
\end{eqnarray}

The last integral, we will need, is lengthy and therefore we will not give it in its full complexity. Happily in Section~\ref{sec:dsf} only singular terms when $p^2\rightarrow m^2$ are necesssary:

\begin{eqnarray}
&&\!\!\!\!\!\int dM\rho(M)\bigg(\frac{1}{m^2-p^2}\times\label{eq:rhoid}\\
&&\times\frac{1}{(m^2x-p^2x(1-x)+\tau^2(1-x))^{1/2}}-(m\rightarrow M)\bigg)\nonumber\\
&&\approx\frac{m^{2\beta}}{(p^2 x^2+\tau^2(1-x))^{1/2}}\frac{1}{(m^2-p^2)^{\beta+1}}\, .\nonumber\\
\nonumber
\end{eqnarray}

\section{Parametric integrals}
\label{sec:pi}

The parametric integrals needed for the vacuum polarization tensor in Section~\ref{sec:dsb} are

\begin{eqnarray}
\int_0^1 dx&&\!\!\!\!\!\frac{x(1-x)}{(1-x(1-x)k^2/m^2)^{\beta+1/2}}\nonumber\\ 
 &&=
\frac{1}{6}\:{}_2F_1(2,\beta+1/2;5/2;k^2/4m^2)\;\; .
\label{eq:in1}
\end{eqnarray}
\begin{eqnarray}
\int_0^1 dx&&\!\!\!\!\!\frac{1}{(1-x(1-x)k^2/m^2)^{\beta+1/2}}\nonumber\\ 
 &&=
\:{}_2F_1(1,\beta+1/2;3/2;k^2/4m^2)\;\; .
\label{eq:in2}
\end{eqnarray}

In Section~\ref{sec:dsf} there appear two other parametric integrals, we denoted by ${\cal S}_1$ and ${\cal S}_2$ (formulae~(\ref{eq:s12a}) and~(\ref{eq:s12b})), which may, after elementary calculation, be expressed by simple functions:  
\begin{eqnarray}
{\cal S}_1&\!\!\!=&\!\!\! m{\cal K}_1(m^2,\tau^2)+2\tau{\cal K}_2(m^2,\tau^2)\label{eq:s1}\\
&\!\!\!=&\!\!\!\int\limits_0^1 dx\frac{m(x+2)+2\tau}{(m^2x^2+\tau^2(1-x))^{1/2}}\nonumber\\
&\!\!\!=&\!\!\!1-\frac{\tau}{m}+\left(2+\frac{2\tau}{m}+\frac{\tau^2}{2m^2}\right)\ln (2m/\tau+1)\; ,\nonumber\\
{\cal S}_2&\!\!\!=&\!\!\!{\cal K}_1(m^2,\tau^2)+2m^2\frac{\partial {\cal K}_1(m^2,\tau^2)}{\partial m^2}+4m\tau\frac{\partial {\cal K}_2(m^2,\tau^2)}{\partial m^2}\nonumber\\
&\!\!\!=&\!\!\!-\tau\int\limits_0^1 dx\frac{(2m+\tau)x^2+\tau(x-2)}{(m^2x^2+\tau^2(1-x))^{1/2}}\nonumber\\
&\!\!\!=&\!\!\! \frac{1}{m}\left[2+\frac{2\tau}{m}-\left(\frac{2\tau}{m}+\frac{\tau^2}{m^2}\right)\ln (2m/\tau+1)\right]\; ,\label{eq:s2}
\end{eqnarray}
where ${\cal K}_1$ and ${\cal K}_2$ were defined in equations~(\ref{eq:k1}) and~(\ref{eq:k2}).


\vfill

\begin{thebibliography}{99}
\bibitem{schw} J. Schwinger, in {\it Theoretical Physics},  
Trieste Lectures 1962 (I.A.E.A., Vienna 1963), p. 89; Phys. Rev. {\bf 128},  
2425(1962).
\bibitem{cks} A. Casher, J. Kogut and L. Susskind, Phys. Rev. 
{\bf D 10}, 732(1974).  
\bibitem{cadam1} C. Adam, Z. Phys. {\bf C 63}, 169(1994).
\bibitem{maie} G. Maiella and F. Schaposnik, Nucl. Phys. {\bf B 132}, 
357(1978).
\bibitem{rot} K.D. Rothe and J.A. Swieca, Ann. Phys. {\bf 117}, 382(1979).
\bibitem{gmc} G. McCartor, Int. J. Mod. Phys. {\bf A 12}, 1091(1997).
\bibitem{trsing} T. Rado\.zycki, Phys. Rev. {\bf D 75}, 085005(2007), Acta Phys. Polon. {\bf B 40}, 1653(2009).
\bibitem{sw} I. O. Stamatescu and T. T. Wu, Nucl. Phys. {\bf B 143},
503(1978). 
\bibitem{os} O. Schnetz, Ph. D. Thesis, Nurnberg 1995.
\bibitem{trpert} T. Rado\.zycki, Eur. Phys. J. {\bf C 6}, 549(1999).
\bibitem{cjs} S. Coleman, R. Jackiw and L. Susskind, Ann. 
Phys. (N.Y.) {\bf 93}, 267(1975). 
\bibitem{ad1} C. Adam, Ann. Phys. (N.Y.) {\bf 259} 1(1997). 
\bibitem{bpr} C. J. Burden, J. Praschifka and C. D. Roberts, Phys. Rev. {\bf D 46} 2695(1992).
\bibitem{maris} P. Maris, Phys. Rev. {\bf D 52}, 6087(1995).
\bibitem{brcr} A. Bashir {\em et al.}, Phys.Rev. {\bf C 78}, 055201(2008).
\bibitem{sauli} V. \u{S}auli, Acta Phys. Polon B Proc. Suppl. {\bf 2}, 443(2009);
\bibitem{srb} S. Sanchez, A. Raya and A. Bashir, AIP Conf.Proc. {\bf 1116} 461(2009).
\bibitem{kog} W. Armour, J. B. Kogut and C. Strouthos, Phys.Rev. {\bf D 82} 014503(2010).
\bibitem{abcw} T. Appelquist {\em et al.}, Phys. Rev. Lett {\bf D 33}, 3704(1986).
\bibitem{abkw} T. Appelquist {\em et al.}, Phys. Rev. {\bf 55}, 1715(1985). 
\bibitem{ab} T. W. Allen and C. J. Burden, Phys. Rev. {\bf D 53}, 5842(1996) ,{\em ibid.} {\bf D 54} 6567(1996).
\bibitem{matna} T. Matsuyama and H. Nagahiro, Bull. Nara. Univ. Educ. {bf 50}, 1(2001).
\bibitem{hosh} Y. Hoshino, arXiv:0706.1603.
\bibitem{bhr} A. Bashir, A Huet and A. Raya, Phys. Rev. {\bf D 66}, 025029(2002).
\bibitem{abar} A. Bashir and A Raya, Few Body Syst. {\bf 41}, 185(2007).
\bibitem{djt} S. Deser, R. Jackiw and Templeton, Ann. Phys. {\bf 140}, 372(1982).
\bibitem{tsvelik} A.M. Tsvelik, {\em Quantum Field Theory in Condensed Matter Physics}, Cambridge Univ. Press, Cambridge 1996.
\bibitem{car} M. S. Carena, T. E. Clark and C. E. M. Wagner, Int. J. Mod. Phys. {\bf A 6} 217(1991).
\bibitem{lyk} J. D. Lykken and J. Sonnenschein, Phys. Rev. {\bf D 42} 2161(1990).
\bibitem{tozha} G. Thompson and R. Zhang, J. Phys {\bf G 13}, L93(1987).
\bibitem{trinst} T. Rado\.zycki, Phys. Rev. {\bf D 60}, 105027(1999).
\bibitem{trjmn} T. Rado\.zycki and J. M. Namys{\l}owski, Phys. Rev. {\bf D 59}, 065010(1999).
\bibitem{bc} J. S. Ball and T.-W. Chiu, Phys. Rev. {\bf D 22}, 2542(1980).
\bibitem{br} C. J. Burden and C. D. Roberts Phys. Rev. {\bf D 44}, 540(1991).
\bibitem{cp} D. C. Curtis, M. R. Pennington, Phys. Rev. {\bf
D 42}, 4165(1990).
\bibitem{kiz} A. Kizilersu and M. R. Pennington, Phys. Rev. {\bf D 79}, 125020(2009).
\bibitem{brsm} A. Bashir, A. Raya and S. S\'anchez-Madrigal, Phys. Rev. {\bf D 84} , 036013(2011).
\bibitem{sal2} A. Salam, Phys. Rev. {\bf 130}, 1287(1963).
\bibitem{del1} R. Delbourgo, N. Cim. {\bf 49 A}, 484(1979).
\bibitem{tz} G. Thompson, R. Zhang, Phys. Rev. {\bf D 35}, 631
(1987).
\bibitem{dz} R. Delbourgo, R. Zhang, J. Phys {\bf A 17}, 3593
(1984). 
\bibitem{park} C. N. Parker, J. Phys. {\bf A 17}, 2873
(1984).
\bibitem{tribb} T. Rado\.zycki and I. Bia{\l}ynicki-Birula, Phys. Rev. {\bf D 52}, 2439(1995).
\bibitem{ch} S. Coleman and B. Hill, Phys. Letters {\bf 159 B}, 184(1985).
\bibitem{wei} S. Weinberg, {\em The Quantum Theory of Fields}, Cambridge University Press, New York 1995.
\bibitem{moroz} S. Moroz, arXiv:0710.4880(2008).
\bibitem{mnu} T. Matsuyama, H. Nagahiro and S. Uchida, Phys. Rev. {\bf D 60}, 105020(1999).
\bibitem{dyson} F. J. Dyson, Phys. Rev. {\bf 85}, 631(1952).
\bibitem{vawi} C. Vafa and E. Witten, Nucl. Phys. {\bf B 234}, 173(1984); Phys. Rev. Lett. {\bf 53}, 535(1984).
\bibitem{ein} M. E. Einhorn and J. Wudka, Phys. Rev. {\bf D 67}, 045004(2003).
\bibitem{tapp} T. Appelquist {\em et al}. , Phys. Rev. {\bf D 33}, 3774(1986).
\end{thebibliography}
\end{document}